\documentclass[preprint,12pt,3p]{elsarticle}

\usepackage{matlab-prettifier}
\usepackage{graphicx}
\usepackage{amsmath}
\usepackage{makecell}
\usepackage{xfrac}
\usepackage{parskip}
\usepackage[backref=page]{hyperref}
\usepackage{adjustbox}
\usepackage{setspace}
\hypersetup{ 
backref=true, 
bookmarks=true, 
pdftoolbar=true, 
pdfmenubar=true, 
pdffitwindow=false, 
pdfstartview={FitH}, 
pdftitle={}, 
pdfauthor={}, 
pdfsubject={}, 
pdfkeywords={}{}, 
pdfnewwindow=true, 
colorlinks=true, 
linkcolor=BlueViolet, 
citecolor=maroon, 
urlcolor=blue,
}
\usepackage{comment}
\usepackage{subfig}
\usepackage{caption}
\usepackage{longtable}
\usepackage{amssymb}
\usepackage{gensymb}
\usepackage{siunitx}

\usepackage{natbib}
\biboptions{comma,round}
\setcitestyle{authoryear}

\begin{document}
\begin{frontmatter}

\title{A meta-analysis of the total economic impact of climate change}

\author[label1,label2,label3,label4,label5,label6]{Richard S.J. Tol\corref{cor1}
}
\address[label1]{Department of Economics, University of Sussex, Falmer, United Kingdom}
\address[label2]{Institute for Environmental Studies, Vrije Universiteit, Amsterdam, The Netherlands}
\address[label3]{Department of Spatial Economics, Vrije Universiteit, Amsterdam, The Netherlands}
\address[label4]{Tinbergen Institute, Amsterdam, The Netherlands}
\address[label5]{CESifo, Munich, Germany}
\address[label6]{Payne Institute for Public Policy, Colorado School of Mines, Golden, CO, USA}

\cortext[cor1]{Jubilee Building, BN1 9SL, UK}

\ead{r.tol@sussex.ac.uk}
\ead[url]{http://www.ae-info.org/ae/Member/Tol\_Richard}

\begin{abstract}
Earlier meta-analyses of the economic impact of climate change are updated with more data, with three new results: (1) The central estimate of the economic impact of global warming is always negative. (2) The confidence interval about the estimates is much wider. (3) Elicitation methods are most pessimistic, econometric studies most optimistic. Two previous results remain: (4) The uncertainty about the impact is skewed towards negative surprises. (5) Poorer countries are much more vulnerable than richer ones. A meta-analysis of the impact of weather shocks reveals that studies, which relate economic growth to temperature levels, cannot agree on the sign of the impact whereas studies, which make economic growth a function of temperature change do agree on the sign but differ an order of magnitude in effect size. The former studies posit that climate change has a permanent effect on economic growth, the latter that the effect is transient. The impact on economic growth implied by studies of the impact of climate change is close to the growth impact estimated as a function of weather shocks. The social cost of carbon shows a similar pattern to the total impact estimates, but with more emphasis on the impacts of moderate warming in the near and medium term.
\end{abstract}

\end{frontmatter}

\section{Introduction}
Estimates of the total economic impact of climate change underpin the social cost of carbon, which determines the optimal rate of greenhouse gas emission reduction. The number of estimates of the total impact has risen rapidly in recent years, so that previous meta-analyses \citep{Howard2017, NordhausMoffat2017, Tol2018} are now out of date. The literature comprises a wide range and seemingly incommensurate estimates of the effects of climate change and weather shocks. This paper reconciles different estimates and updates previous meta-analyses.

Studies of the economic impact of climate change use a range of methods\textemdash enumeration, elicitation, computable general equilibrium, econometrics\textemdash each with its pros and cons. Studies of the economic impact of weather are exclusively econometric but use two alternative specifications, one in which \emph{temperature} affects economic growth and one in which \emph{temperature change} affects growth. As shown below, the former is inconsistent with the climate literature. The latter is consistent in principle, conditional on a scenario and model.

This paper makes three contributions. First, I update my earlier meta-analysis of the economic impact of climate change \citep{Tol2009JEP, Tol2014JEP, Tol2018}, using the same methods as before. This confirms some earlier findings but overturns others. In contrast to \citet{Howard2017}, I separately analyze weather and climate estimates. I use more estimates than they do, but skip some of their numbers. They appear to have misread some papers, included some estimates twice, and added estimates that are not. See \ref{sc:howard} for details. Compared to \citet{NordhausMoffat2017}, I add more estimates, avoid arbitrary weights, and include estimates of the impact of weather shocks. I use more estimates than \citet[][see also \citet{Tol2016}]{Rose2022}, and separate the impacts of climate change and weather shocks.

As a second contribution, I present the first meta-analyses of the two literatures on the economic impact of weather shocks. \citet{auffhammer2018quantifying} and \citet{Kolstad2020} discuss this literature but do not combine estimates. Third, I propose a method to reconcile one of the weather literatures with the climate literature.

The paper proceeds as follows. Section \ref{sc:climate} discusses the literature on the estimated impacts of climate change. This section follows \citet{Tol2021anyas}, but with new numbers. Section \ref{sc:weather} turns to the impacts of weather shocks. Section \ref{sc:reconciliation} reconciles these estimates where possible. Section \ref{sc:scc} shows the implications for the social cost of carbon. Section \ref{sc:conclusion} concludes.

\section{Impacts of climate change}
\label{sc:climate}
Table \ref{tab:compstat} shows 61 estimates, from 33 studies, of the total economic impact of climate change. These estimates are comparative static, comparing economies of the recent past with and without some future climate change. Figure \ref{fig:totalimpact} plots 58 of the estimates; Figure \ref{fig:totalimpactall} plots them all, including 3 estimates for very large warming. The horizontal axis is the increase in the global annual mean surface air temperature. The vertical axis is the welfare-equivalent income change, some approximation of the Hicksian Equivalent Variation. These numbers should be read as follows: A global warming of 2.5\celsius{} would make the average person feel as if she had lost 1.7\% of her income. 1.7\% is the average of the 13 dots at this level of warming.

\subsection{Methods}
As indicated in Figure \ref{fig:totalimpact} and Table \ref{tab:compstat}, these estimates use a range of methods. Older studies tend to rely on the enumerative method. Estimates of the impacts (after assumed adaptation) of climate change in their natural units are multiplied with estimates of their values and added up. This is a direct cost, a poor approximation of the change in welfare. The enumerative approach omits price changes and interactions between sectors, such as changes in water resources affecting agriculture.

Price changes and market interactions between sectors are included in estimates that use a computable general equilibrium model (CGE). This method has become more popular over the years. These studies report the Hicksian Equivalent Variation but, as they are based on the national accounts, omit direct welfare impacts on health and nature.\footnote{One CGE study \citep{Takakura2019} uses the value of a statistical life to assess health impacts.} CGE models allow for adaptation in the production function\textemdash for example, a drop in the productivity of land would be partially compensated by an increase in the application of fertilizer and labour\textemdash and through shifts in supply and demand.\footnote{If adaptation is implied in the assumed shock, then there is no adaptation in production. For example, many CGEs take their impact of climate change on labour productivity, one of the largest impacts, from \citet{Kjellstrom2009}, but as temperature is not an input factor, air conditioning is kept constant just as it is in Kjellstrom.}

Econometric studies have also become more numerous over time. Differences in prices, expenditures, self-reported happiness or total output are ascribed to variations in climate. A key advantage of this method is that adaptation is observed rather than assumed, but identification comes from the cross-section and projected impacts from substituting time for space. 

Four studies elicit expert views. Two of these studies were done before anyone could reasonably claim expertise on the economic impacts of climate change, the later ones use a mix of people who have and have not published on the subject. Views were expressed about the impact of climate change on global output, which can be interpreted as a measure of economic activity (but not welfare) as well as a measure of income (and thus welfare).

\subsection{Combining estimates}
Figure \ref{fig:totalimpact} shows a curve, based on the method of \citet{Tol2019EE}. Seven alternative impact functions, proposed in the literature, were fitted to the data. See Table \ref{tab:compstat}.\footnote{I dropped the double-exponential model of \citet{Golosov2014}, as its fit to the data is so poorly, and replaced it with the hyperbolic sine.} I minimized the weighted sum of squared differences between the model and the primary estimate, weighting all estimates from a single study equally and attaching a total weight of 1 per study. The curve shown is the Bayesian model average, that is, the seven impact functions are weighted according to their fit to the primary estimates.

The 90\% confidence interval shown is based on the uncertainty reported in 14 of the primary estimates, following \citet[][see also \citet{Tol2012ERE, Tol2015CE}]{Tol2018}. The lower and upper bound is estimated separately as a linear function of the temperature increase, using weighted least squares as above.

\subsection{Results}
Compared to my previous meta-analysis based on the same methods \citep{Tol2019}, Figure \ref{fig:totalimpact} shows a very different picture. The number of estimates has more than doubled. The number of estimates beyond 3.2\celsius{} of warming has increased tenfold. As there is a non-negligible chance of large warming, the previous paucity of evidence allowed for speculation about the expected impact of climate change \citep{Weitzman2009, Anthoff2022}.

The central estimate of the impact of global warming is always negative, but the confidence interval is too wide to put much confidence in that. The wider confidence interval, compared to \citet{Tol2019}, is mostly due to the range of uncertainties reported by \citet{Howard2020, Howard2021}. The central estimate is higher because of the positive impacts reported by \citet{Desmet2015} and \citet{Newell2021} for substantial global warming. Enumerative studies report positive impacts of climate change due to reduced costs of heating in winter, lower cold-related mortality, and carbon dioxide fertilization. In \citet{Desmet2015}, the positive impacts are in manufacturing.\footnote{The "into Siberia" series of papers \citep{Desmet2015, Conte2021, Cruz2021} assume mobility of capital and labour.} There is no sectoral breakdown in \citet{Newell2021}; the positive impacts are in specifications that relate economic growth to the temperature level (see below).

Researchers disagree on the sign of the net impact, but agree on the order of magnitude: The welfare loss (or gain) caused by climate change is equivalent to the welfare loss caused by an income drop of at most ten percent\textemdash a century of climate change is not worse than losing a decade of economic growth.

The uncertainty is large and right-skewed. For every degree warming, the positive standard deviation increases by 1.02\% GDP while the negative standard deviation increases by 1.43\% GDP. That is, negative surprises are larger than positive surprises of equal probability.

Figure \ref{fig:totalimpact} suggests that different methods yield different results. Figure \ref{fig:method} shows the curve fitted separately by method used for the primary impact estimate. Instead of using the Bayesian average, the curve with the best fit is shown because there are not enough observations to estimate so many parameters if the sample is split into four. The elicitation studies are most pessimistic, the econometric studies most optimistic about the impacts of climate change. The enumerative and general equilibrium papers lie in between, with the former more pessimistic for moderate warming and less pessimistic for more profound warming. Although the central estimates are different, the uncertainty is so large that differences do not become statistically significant from zero before 4\celsius{} of warming.

\subsection{Sectoral impacts and omission bias}
Eight studies show sectoral impacts in tabular format.\footnote{\citet{Takakura2019} show sectoral impacts in hard-to-read graphs.} The results are shown in Table \ref{tab:sector} for 2.5\celsius{} global warming. The estimates by \citet{Sartori2016} and \citet{Kompas2018} are scaled using the function shown in Figure \ref{fig:totalimpact}.

The different studies have different sectoral cover, which I mapped to the sectors shown in Table \ref{tab:sector}. Following \citet{Tol2019HB}, where a particular study omits a certain sector, I impute its value with the average of the studies that do include this impact.

The following results emerge. ``Other markets'' is the biggest impact. Although recorded in an obscure way, this is primarily the impact of heat on labour productivity. Health impacts come second, followed by amenity, extreme weather, and agriculture. Only the impact of climate change on time use is positive. Overall, market impacts make up 55\% of the total, while the remaining 45\% directly affect welfare.

On average, imputation of missing impacts increases the total impact estimate by 63\%. The study by \citet{Tol1995} is most complete, \citet{Sartori2016} least. The estimates in Figure \ref{fig:totalimpact} therefore appear to be underestimates of the true impact.

\subsection{Distribution of impacts}
Figure \ref{fig:totalimpact} shows the global average impact of climate change. 18 of the 33 studies include estimates of the \textit{regional} impacts of climate change or \textit{national} impact estimates. Following \citet{Tol2021}, I regress the estimated regional impact on per capita income and average annual temperature, with dummies $\alpha_s$ for the studies. This yields
\begin{equation}
\label{eq:natimp}
 I_c = \alpha_s + 1.87 (0.52) \ln y_c - 0.39 (0.09) T_c
\end{equation}
where $I_c$ is the impact in country $c$ (in \%GDP), $y_c$ is its average income (in 2010 market exchange dollars per person per year), and $T_c$ is the average annual temperature (in degrees Celsius); the bracketed numbers are standard errors. Hotter countries have more negative impacts. Richer countries have relatively less negative impacts.\footnote{\citet{Sterner2008} and \citet{Bremer2021} assume the opposite, empirical support to the contrary notwithstanding \citep[e.g.][]{Botzen2021, Gandhi2022}.} For each of the studies, this equation is used to impute national impacts, making sure that the regional or global totals match those in the original estimates. The function shown in Figure \ref{fig:totalimpact} is then used to shift all impacts to 2.5\celsius warming.\footnote{Note that I use the \textit{global} function to shift the imputed \textit{national} impacts. \citep{Tol2019} shows that this is more robust than estimating \textit{national} impact functions.} For each country, the average and standard deviation across studies is taken.

\begin{figure}
\centering
\includegraphics[width=\textwidth]{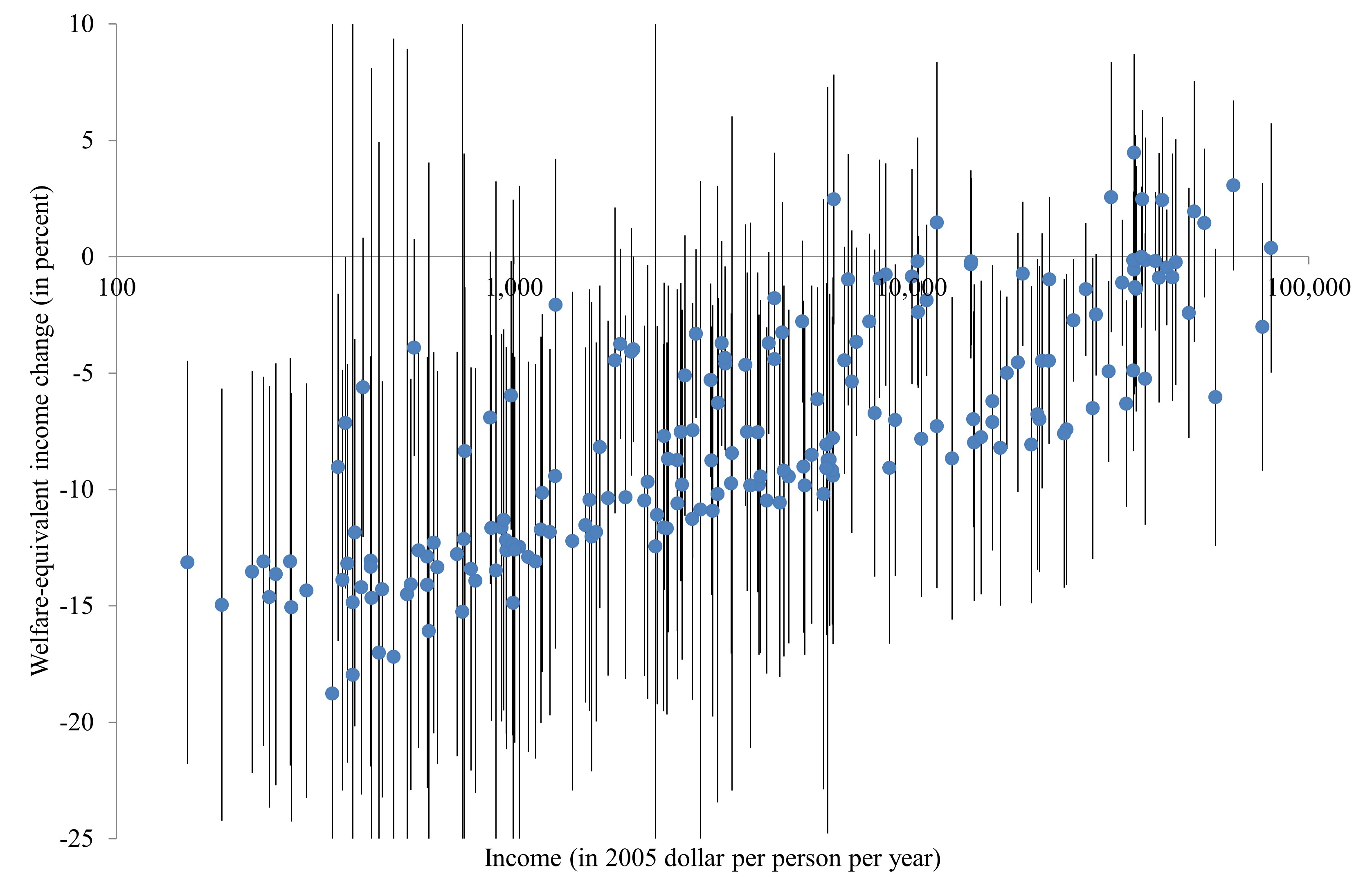}
\includegraphics[width=\textwidth]{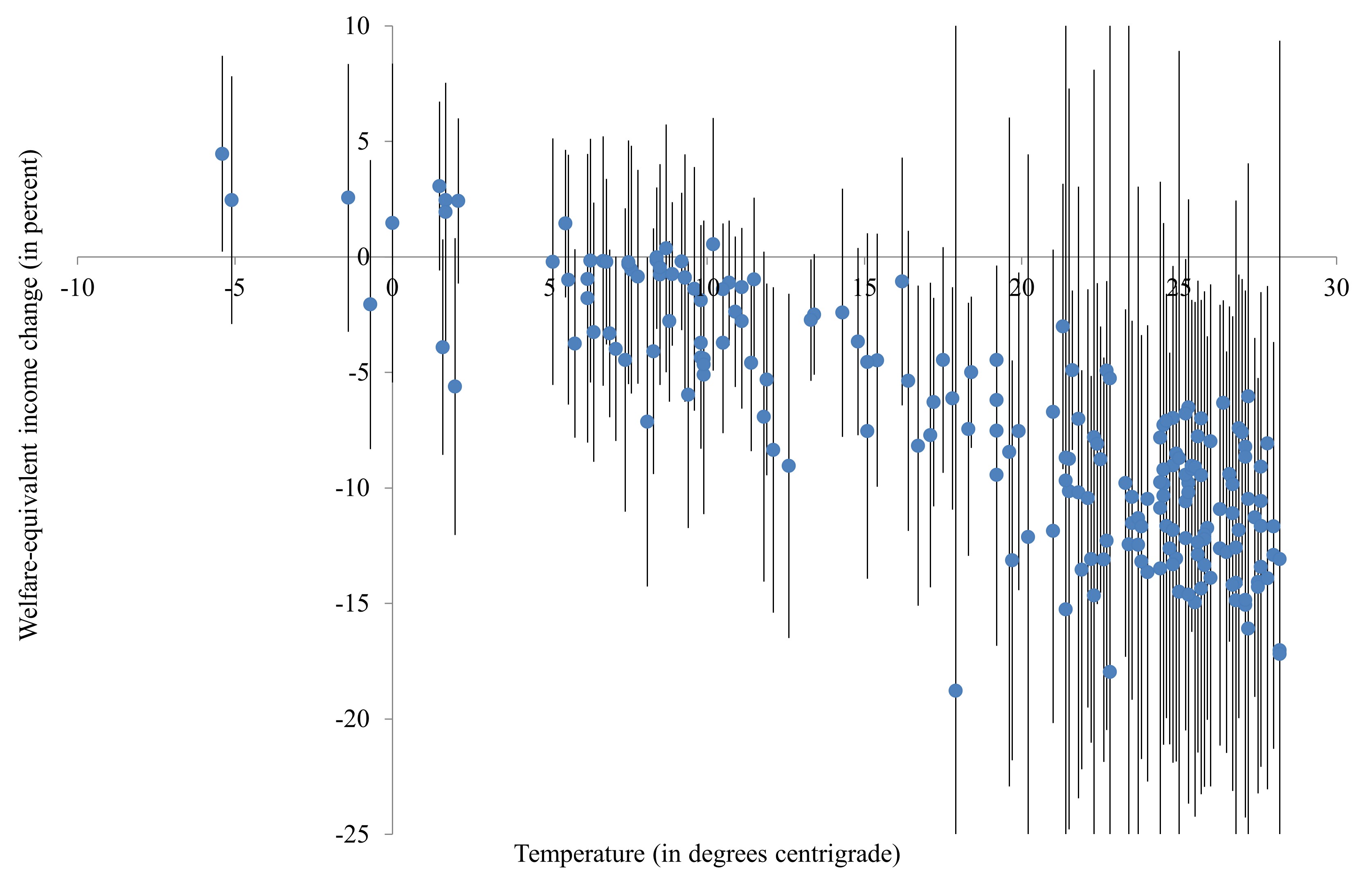}
\caption{The economic impact of climate change for a 2.5\celsius{} warming for countries as a function of their income (top panel) and temperature (bottom panel).}
\label{fig:natimp}
\end{figure}

Figure \ref{fig:natimp} shows results for individual countries for 2.5\celsius{} warming. Hotter countries, poorer countries see more negative impacts. In fact, the majority of countries show a larger damage than the global average of 1.7\%. This is because the world economy is concentrated in a few, rich countries. The world average economic impact counts dollars, rather than countries, let alone people.

Poorer countries are more vulnerable to climate change for three reasons. First, poorer countries have a higher share of their economic activity in sectors, such as agriculture, that are directly exposed to the vagaries of weather. Second, poorer countries tend to be in hotter places. This makes adaptation more difficult as there are no analogues for human behaviour and technology. Cities in temperate climates need to look at subtropical cities to discover how to cope in a warmer climate, and subtropical cities at tropical ones. The hottest cities will need to invent, from scratch, how to deal with greater heat. Third, poorer countries tend to have a limited \textit{adaptive capacity} \citep{ADGER2006, Yohe2002}. Adaptive capacity depends on a range of factors, such as the availability of technology, the ability to pay for those technologies, the political will to mobilize resources for the public good, and the government's competence in raising funds and delivering projects. All these factors are worse in developing countries.

\section{Impacts of weather shocks}
\label{sc:weather}
Climate, the thirty-year average of weather, varies only slowly over time and has not varied much over the period for which data are good. In the econometric studies discussed above, the impact of climate is identified from cross-sectional variation. Many other things vary over space too. Panel data help, but some confounders do not change much over time. Therefore, some researchers have estimated the impacts of \emph{weather} on a range of economic activities. From an economic perspective, weather is random and its impact therefore properly identified. Unfortunately, the impact of a weather shock is not the same as the impact of climate change \citep{Dell2014}. Particularly, weather studies estimate the short-run response of the economy, whereas the interest is in the long-run response, with adjustments in capital, behaviour and technology. \citet{Deryugina2017} derive the rather restrictive conditions under which weather variability is informative about climate change. These conditions are roughly the same as for a market equilibrium to be a Pareto optimum. These conditions are not met. Food markets are distorted by subsidies and import tariffs. Coastal protection is a public good. Infectious disease is an externality. Irrigation is a lumpy investment. \citet{Lemoine2018} notes that economic agents would need to have rational expectations of future weather for investments in adaptation to be optimal. There is little evidence to support that. Extrapolating the impact of weather shocks therefore does not lead to credible estimates of the impact of climate change.

These caveats notwithstanding, there are a number of papers that estimate the economic impact of weather shocks. I restrict the attention to studies of the impact of temperature shocks on economic growth. Different studies use different specifications, but there are two broad clusters. The first, older cluster regresses the growth rate of economic output on temperature \emph{levels}. The second, younger cluster regresses the economic growth rate on the \emph{change} in temperature.\footnote{Rainfall is typically included too, but is often found to be insignificant, except in \citet{Kotz2022}.} I separately discuss these clusters.

\subsection{Temperature levels and economic growth}
Six studies estimate the impact of temperature \emph{levels} on economic growth: \citet{Dell2012}, \citet{Burke2015}, \citet{Pretis2018}, \citet{Henseler2019}, \citet{Acevedo2020} and \citet{Kikstra2021}.\footnote{A seventh study does not report parameter estimates \citep{Callahan2022}.} The impact of temperature levels on economic growth numerically dominates the impact of temperature changes in \citet{Kalkuhl2020}.

Note that all these papers use country fixed effects, nullifying a systematic effect of climate on economic growth rates in sample. However, in this specification, the impact of climate change on economic growth will last forever.

Temperature has a linear impact on economic growth in the preferred specifications of \citet{Dell2012} and \citet{Kalkuhl2020}, but in the former study the impact is significant only in poorer countries. In the other four papers, there is a parabolic relationship between temperature and growth.

\begin{figure}[h]
\includegraphics[width=\textwidth]{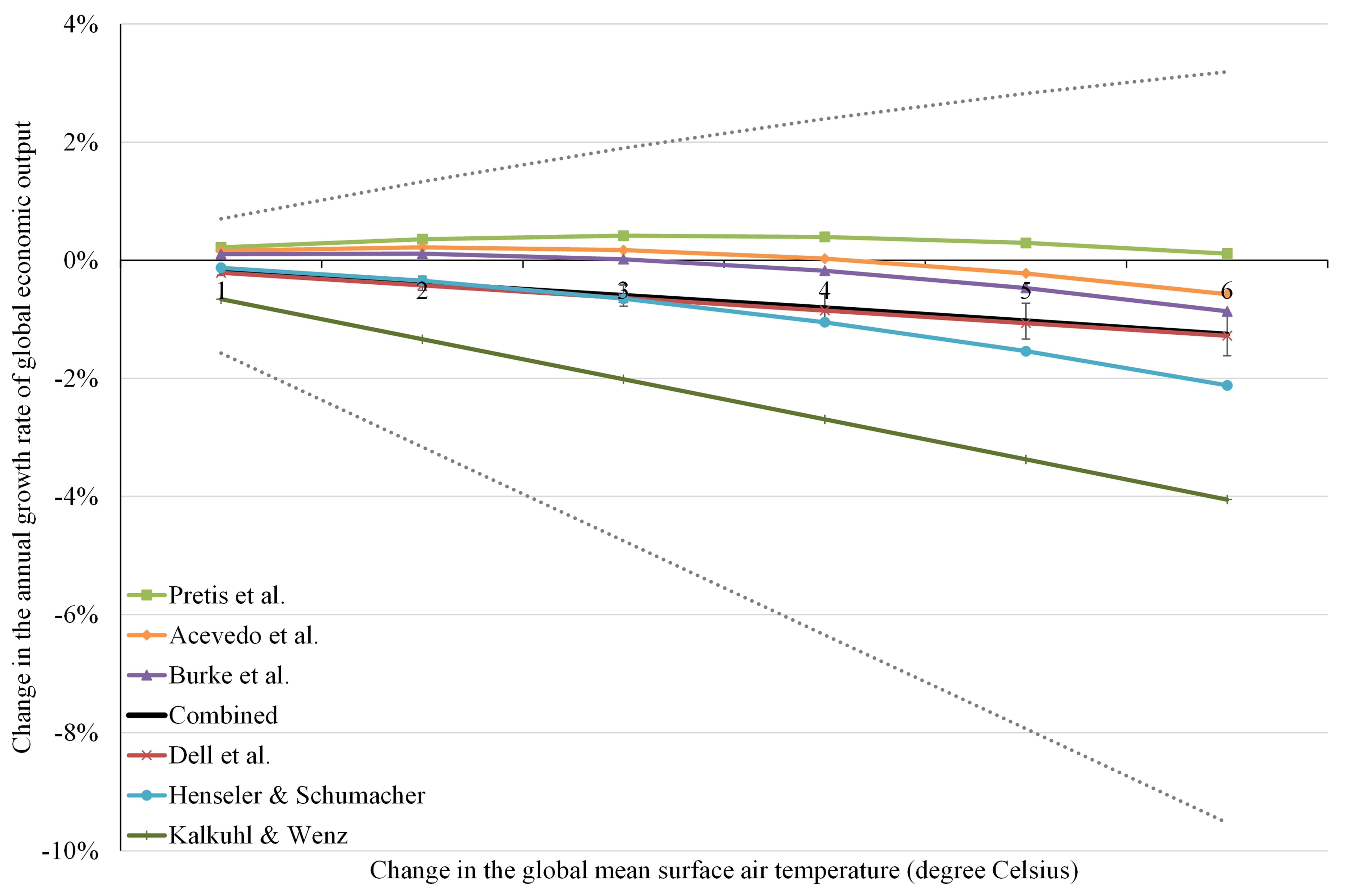}
\caption{The impact of \emph{climate} on global economic growth. Primary estimates are shown in colour. The black line is the shrunk model average, plus or minus its standard error, the dashed lines the maximum (minimum) estimate plus (minus) its standard error.}
\label{fig:growth}
\end{figure}

Figure \ref{fig:growth} shows the effect sizes. The functions are shifted so that, for each country, pre-industrial temperatures have no effect on growth. The global effect is the weighted average of the national effects, using 2015 GDP as weights. Global effect sizes vary between small positive (in three of six studies for moderate warming) and large negative impacts\textemdash \citet{Newell2021} find that only models with positive impacts are supported by cross-validation tests. According to the most pessimistic study \citep{Kalkuhl2020}, 3\celsius{} warming would end economic growth.

Figure \ref{fig:growth} also shows the impact of all six studies together, shrunk to their average.\footnote{Alternatively, use Dell as the prior and update this with the other five studies to find the posterior.} The combined impact is very close to that of the first study \citep{Dell2012}. The confidence interval is narrow. Shrinkage tends to lead to overconfidence and may be inappropriate in this case as these studies use much the same data\textemdash the different effect sizes are therefore all the more striking. Figure \ref{fig:growth} also shows the most optimistic estimate plus its standard error and the more pessimistic one minus its standard error. The resulting interval is wide. This is appropriate as the differences between the central estimates are large too.  

\subsection{Temperature change and economic growth}
Four studies estimate the effect of temperature \emph{change} on economic growth: \citet{Letta2018}, \citet{Kahn2019}, \citet{Kotz2021} and \citet{Kotz2022}. As noted above, \citet{Kalkuhl2020} also estimate this but prefer a specification in which the impact of temperature level dominates the impact of temperature change.

\citet{Letta2018} finds a linear effect that only affects poor countries. \citet{Kotz2021,Kotz2022} also use a linear specification but weather shocks affect all countries; the effect is statistically significantly larger in warmer countries, but the effect size is too small to meaningfully affect projections of climate change. \citet{Kahn2019} find that all countries are affected equally, but that hot shocks have a larger impact than cold shocks.

Weather is stochastic. In a stationary climate, the models of \citet{Letta2018} and \citet{Kotz2021, Kotz2022} predict that the impact of negative and positive temperature shocks cancel out. In a warming climate, positive shocks are more likely and, as the models are linear, the expected impact on growth is proportional to the rate of warming. In the model of \citet{Kahn2019}, a stationary climate does have a negative effect on economic growth, assumed to be offset by the country fixed effects.\footnote{As they define temperature shocks relative to the mean, the impact of climate change is transient by construction, just as it is in \citet{Letta2018} and \citet{Kotz2021, Kotz2022}, provided that climate change does not affect the skewness of the temperature distribution.} In a warming climate, the expected impact on growth is the sum of average reduction in negative shocks and the average increase in positive shocks.

\begin{figure}[h]
\includegraphics[width=\textwidth]{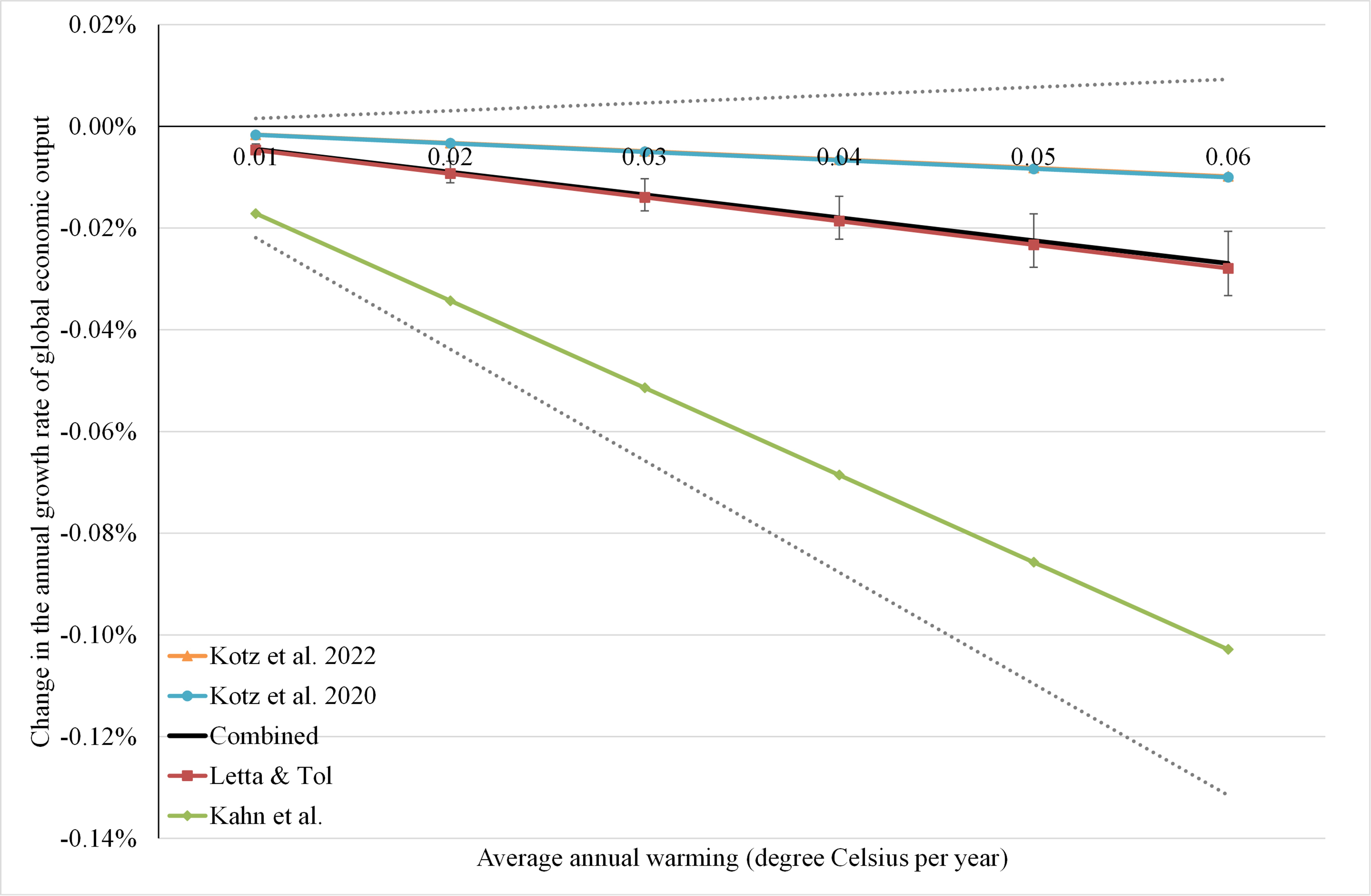}
\caption{The impact of \emph{climate change} on global economic growth. Primary estimates are shown in colour. The black line is the shrunk model average, plus or minus its standard error, the dashed lines the maximum (minimum) estimate plus (minus) its standard error.}
\label{fig:level}
\end{figure}

Figure \ref{fig:level} shows the estimated effect sizes, as a function of the rate of warming rather than its level. The two results by \citet{Kotz2021, Kotz2022} are very similar and least pessimistic. \citet{Kahn2019} shows the largest effects. \citet{Letta2018} is somewhere in between and very close to the combined effect. Shrinkage leads again to overly confident results, so Figure \ref{fig:level} also shows the top of Kotz' confidence interval and the bottom of Kahn's.

\section{Reconciling estimates}
\label{sc:reconciliation}
There are two sets of estimates of the economic impact of climate change. The first set \emph{directly} studies the impact of climate change. A range of methods is used, including enumerative studies, computable general equilibrium models, Ricardian or other econometric techniques, and elicitation methods. The second set studies the impact of weather shocks, exclusively using econometric methods, and from this one \emph{infers} the effects of climate change. Other key differences between the two set of studies is that the former leads to a \emph{level} effect on \emph{welfare} while the latter is a \emph{growth} effect on \emph{output}.

Welfare and output effects are easily reconciled if we assume that consumption is proportional to output\textemdash \citet{Pizer1999} shows that the impact of climate change on the savings' rate is minimal\textemdash and note that output studies are incomplete for ignoring the intangible losses due to climate change. For example, some studies of the impact of climate and weather on output assume that health effects are adequately captured by expenditure on medical and funeral services.

A further complication is that welfare effects may arise from a change in the \emph{composition} rather than the \emph{level} of economic output. Defensive expenditure\textemdash additional coastal protection due to sea level rise, for instance\textemdash counts \emph{towards} economic output but \emph{against} welfare. \citet{Tol1998} estimate that 7-25\% of impact estimates are the costs of adaptation; I use their midpoint of 16\% below. Again, estimates of the impact of weather shocks on economic output are a lower bound on estimates of the impact on welfare.

Level and growth effects are harder to reconcile. Previous studies mixed growth and level effects without further ado in a graph \citep{Kahn2019, Rose2022} or a meta-analysis \cite{Howard2017}. This is inadequate. A growth effect implies a level effect, of course, but for a particular year, and conditional on assumptions on the accumulation of weather shocks into climate change and on the shape of the impact function. \citet{Fankhauser2005} show that the growth effect implied by a level effect is contingent on the assumed growth model and its parameters as well as on the rate of climate change. Below, I assess the growth effect of the level impact. I then accumulate both implied and estimated growth effects along a particular scenario so as to compare the results of these two strands of literature.

\subsection{Growth effects}
The economic impact of climate change has an apparent effect on economic growth. Suppose that climate change scales down economic output by a factor $\frac{1}{1+D_t}$, where $D$ denotes the economic impact of climate change \citep{Nordhaus1992, Nordhaus1993}. If output is a Cobb-Douglas function of capital and labour then, by log-linearization and differentiation, the growth rate of the economy
\begin{equation}
 \dot{Y}_t \approx \dot{A}_t + \lambda \dot{K}_t + (1-\lambda) \dot{L}_t - \dot{(1+D_t)}
\end{equation}
where $Y_t$ is output at time $t$, $A$ is total factor productivity, $K$ is the capital stock, $L$ is the labour force and $\lambda$ is the capital elasticity of output; $\dot{Y}_t = \frac{\partial Y_t}{\partial t}$. Growth is slower if $D_t>D_{t-1}>0$, that is, if climate change damages are increasing.

The \emph{apparent} impact on economic growth is probably small. If the impact is 1.00\% in year $t$ and 1.01\% in the year after, $\dot{(1+D_t)} = 0.01\%$. The apparent growth effect is large only if damages are high, if the impact function is highly non-linear, or if climate change very rapid.

The \emph{actual} impact is larger than the apparent one. Contrast two output levels
\begin{equation}
 Y_t = \frac{A_t K_t^\lambda L_t^{1-\lambda}}{1+D_t}
\end{equation}
for $D=0$ and $D>0$.

In a Solow-Swan model of economic growth, $A$ and $L$ are exogenous. Then the change in the growth rate of the economy
\begin{equation}
\label{eq:growth0}
 \Delta \dot{Y}_{t+1} := \dot{Y}_{t+1|D=0} - \dot{Y}_{t+1|D>0} = \lambda \frac{s \left ( Y_t - \frac{Y_t}{1+D_t} \right ) }{K_t} = \lambda \frac{sY_t}{K_t} \frac{D_t}{1+D_t} = \lambda \delta \frac{D_t}{1+D_t} 
\end{equation}
where $s$ is the savings rate. The final equality holds if the capital stock is in its steady-state $K=\frac{s}{\delta}Y$. Equation (\ref{eq:growth0}) holds for the year immediately following the impact. In later years, the impact on growth equals
\begin{equation}
\label{eq:growth}
 \Delta \dot{Y}_{t+\tau} := (\lambda \delta)^\tau \frac{D_t}{1+D_t} 
\end{equation}
This shows that the impact of climate change on economic growth is transient: The effect fades away\textemdash as in the econometric studies that have growth depend on the change in temperature.

The growth effect fades away quickly. $\lambda \cdot \delta = 0.3 \cdot 0.1 = 0.03$ would be reasonable parameter choices. The cumulative effect is around: $\sum_{\tau=1}^{\infty} (\lambda \delta)^\tau \approx 0.031$.

In new growth models, where total factor productivity is endogenous, the impact of an output shock is also transient. In unified growth models, where the labour force too is endogenous, shocks are transient too, unless the shock happens to push the economy from a Solowian to a Malthusian equilibrium or vice versa.

The hypothesis, furthered by \citet{Dell2012} and \citet{Burke2015}, that the level of climate change would affect the growth rate of the economy is thus inconsistent with the theory of economic growth. \citet[][Appendix A2]{Kahn2019} reach the same conclusion and discuss the implied econometric issues. The Dell-Burke hypothesis is tested by \citet{Kotz2021, Kotz2022} and rejected. \citet{Kalkuhl2020} similarly estimate a model in which both temperature \emph{level} and temperature \emph{change} affect economic growth, to find that the level is not significant; they nonetheless keep levels in their preferred specification, and this effect dominates numerically. \citet{Newell2021} use cross-validation tests; out of 800 specifications, only 9 are valid, 7 where temperature change affects economic growth and 2 where temperature level affects economic growth.\footnote{These 9 valid specification are included in Figure \ref{fig:totalimpact}.}

\subsection{Results}
\begin{figure}[h]
\includegraphics[width=\textwidth]{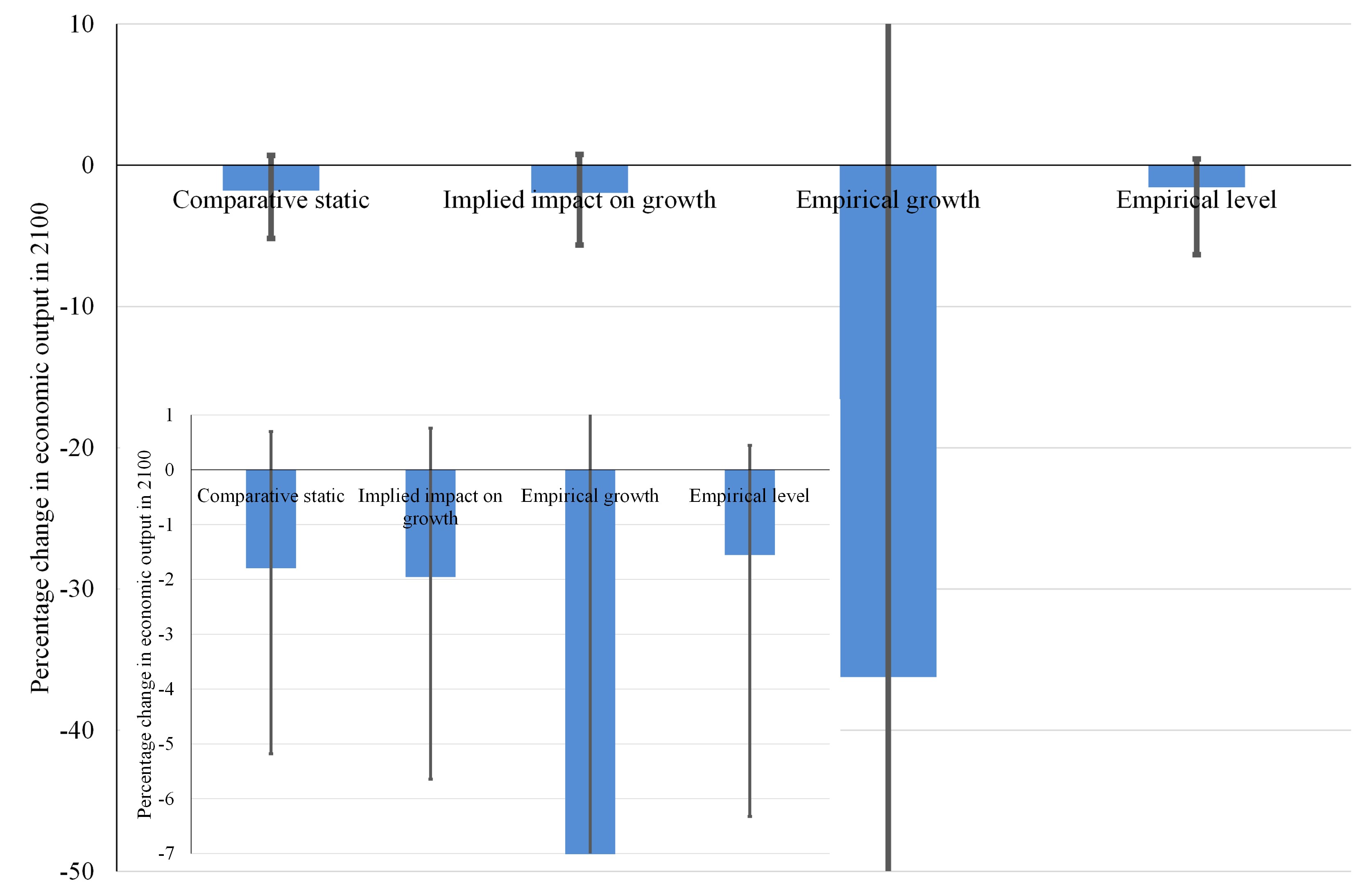}
\caption{The impact of 4.3\celsius{} warming on global economic output in 2100 according to, from left to right, comparative static studies, the growth effect implied by comparative static studies, econometric studies of the impact of climate on growth, and econometric studies of the impact of climate change on growth. The inset shows the same information but compresses the vertical axis.}
\label{fig:compare}
\end{figure}

Figure \ref{fig:compare} compares the above results. I assume that the world was 1.1\celsius{} warmer in 2020 than in the time just before the start of the industrial revolution \citep{Gulev2021}. Following \citet{Newell2021}, I assume that the world will warm on average 0.04\celsius{} per year to reach 4.3\celsius{} by 2100.\footnote{This is not very likely \citet{Srikrishnan2022}, but makes comparison to previous results easier.}

The first column in Figure \ref{fig:compare} shows the comparative static impact as discussed in Section \ref{sc:climate} for 4.3\celsius{} warming, limited to the \emph{market} impacts and assuming that 39\% of impacts lead to a \emph{reduction} of economic output.\footnote{55\% of impacts are market impacts, 16\% are defensive investment.} The second column shows the implied impact on economic growth, accumulated over the 80 year period. The latter effect is somewhat smaller than the former, 1.80\% v 1.65\% of GDP.

The third column shows the cumulative effect on economic output according to the econometric studies that assume that the level of temperature affects economic growth. These studies are very pessimistic: Economic output would be 36\% smaller with than it would be without climate change.

The fourth and final column shows the cumulative effect for the studies that assume that the rate of warming affects economic growth. The central estimate is 1.55\% of GDP. This is somewhat smaller than the growth effect estimated above.

Figure \ref{fig:compare} also includes the indicative 67\% confidence intervals. The comparative static studies show the widest range, the empirical level studies the narrowest. The empirical growth studies are at odds with the rest of the literature. The empirical level studies and the comparative static ones by and large agree.

\section{Implications for the social cost of carbon}
\label{sc:scc}
The social cost of carbon, or its marginal damage, depends on the total impact of climate change. It also depends on the emissions scenario, the parameterisation of the carbon cycle, the rate and extent of warming, and the aggregation of impacts across people, between scenarios, and over time. Instead of exploring that vast parameter space, I report a limited analysis:
\begin{itemize}
    \item For each of the 69 estimates in Table \ref{tab:compstat}, I fit Nordhaus' impact function, the more common among the four single-parameter ones in Table \ref{tab:function}, and compute the social cost of carbon.
    \item I compute the social cost of carbon for the central line in Figure \ref{fig:totalimpact} and the four graphs in Figure \ref{fig:method}.
    \item I compute the social cost of carbon for the graphs in Figures \ref{fig:growth} and \ref{fig:level}.
    \item I use a single scenario (SSP2), a single carbon cycle, a single climate sensitivity (3\celsius/tCO\textsubscript{2}), and a single discount rate (a Ramsey rate with a 1\% pure rate of time preference and a relative rate of risk aversion of 1). These are the central values in \citet{Tol2019}.
    \item I assume away uncertainty and ambiguity. There is no disaggregation of impacts so that inequity aversion is irrelevant.
\end{itemize}
This makes for a total of 95 estimates of the social cost of carbon.\footnote{Note that Equation (\ref{eq:natimp}) has that per capita income affects total but not marginal impacts.}

Table \ref{tab:compstat} shows the estimated social cost of carbon for the 69 estimates of the impact of climate change. The social cost of carbon ranges from -\$355/tC to +\$587/tC. The lower bound is due to \citet{Tol2002ERE1}, who finds positive impacts in the near-term; by construction, impacts are then always positive. The upper bound is due to \citet{Horowitz2009}, who finds large negative impacts in the near-term. The weighted average is \$59/tC, an estimate that is well in line with the literature \citep{Tol2021scc}.

Table \ref{tab:function} shows the estimated social cost of carbon for the 8 alternative impact functions, fitted to all 69 estimates of the total impact. The social cost of carbon varies between \$3/tC for the exponential function and \$29/tC for the linear function. While the exponential function projects the highest total impacts in the distant future, the linear function projects substantial impacts in the near future. The social cost of carbon is discounted, and thus highest under linearity \citep{Peck1994}. The higher estimates are a better fit to the data so that the social cost of carbon for the Bayesian model average is \$27/tC.

Table \ref{tab:methods} shows the estimated social cost of carbon for the four alternative estimation methods, using the best-fitting impact function. Confirming Figure \ref{fig:method}. the elicitation studies are the most pessimistic with an estimate of \$87/tC. The econometric studies, which are essentially undecided about the impact until it gets really hot, are most optimistic with a social costs of carbon of \$1/tC.

Studies of the impact of weather shocks that relate economic growth to temperature are specific to each country. The global impact is approximately quadratic. See Figure \ref{fig:level}. Table \ref{tab:level} shows the fitted parameters, as well as the estimated social cost of carbon, which ranges from -\$171/tC for the study of \citet{Pretis2018} to +\$539/tC for \citet{Kikstra2021}, reflecting the respective impacts in the near- and mid-term. Shrinking the 7 alternative estimates leads to a social cost of carbon of \$218/tC, which is high compared to the estimates based on the impact of climate change.

Table \ref{tab:change} shows the parameters and the social cost of carbon of the weather studies that relate growth to temperature change. Estimates range from \$16/tC for \citet{Letta2018} to \$60/tC for \citet{Kahn2019}. The combined estimate is \$17/tC, close to three of the four studies, and somewhat lower than the climate impact estimates.

Unsurprisingly, the results for the social costs of carbon are in line with the results for the total impact of climate change. There is a wide range of uncertainty. Different methods yield different results. Elicitation leads to pessimistic estimates, and relating economic growth to temperature levels appears to be unreliable.

\section{Discussion and conclusion}
\label{sc:conclusion}
I do four things in this paper. First, I update my earlier meta-analysis using the same methods but much more data. Three new results emerge from this. (1) The initially positive impacts have vanished: The central estimate of the economic impact of global warming is always negative. (2) The confidence interval about the estimates has widened considerably, and now includes \emph{no impact} for very considerable warming. The more we learn about the economic impacts of climate change, the better we understand our ignorance. (3) Elicitation methods give the most pessimistic results, econometric studies the most optimistic ones, with the enumerative methods and computable general equilibrium models in the middle. Two old results remain. (4) The uncertainty about the impact is skewed towards negative surprises. (5) Poorer countries are much more vulnerable than richer ones.

The second contribution is a meta-analysis of the impact of weather shocks, in two parts. Studies that relate economic growth to temperature levels cannot agree on the sign of the impact. Studies that make economic growth a function of temperature change differ an order of magnitude in effect size, but do agree on the sign. The former studies posit that climate change has a permanent effect on economic growth, the latter that the impact is transient.

The third contribution reconciles climate change and weather shocks. The impact on economic growth implied by studies of the impact of climate change are close to the growth impact estimated as a function of temperature change. Growth impacts implied by the temperature level are much larger.

Finally, I assess the implications for the social cost of carbon. The pattern of results for the marginal impact is roughly the same as for the total impact. Because of discounting, the impact of moderate warming is more important for the social cost of carbon than the impact of more pronounced climate change.

The following research gaps appear. Enumerative studies have not been published for a while. New research would reveal whether these papers are really out of date, as is sometimes claimed \citep{NAS2017}. Elicitation studies tend to be pessimistic. It is not clear why supposed experts deviate from the published literature. Computable general equilibrium models draw from the same or similar sets of calibrated impacts, yet produce a range of different impacts. Systematic model comparison would be useful.

The econometric studies, however, show the widest range of results. These come in three groups\textemdash the impact of climate on income, the impact of weather on income growth, and the impact of changes in weather on income growth.\footnote{A survey of the many sectoral and regional climate and weather studies is beyond the current paper.} There is considerably variation within groups, even when data used are much the same, and more variation between groups, both numerically and conceptually. Econometricians are adept at testing which specification fits the data best; these methods should be applied here \citep[][set an example]{Newell2021}. Analyses should be better guided by theory. Data sets should be extended to greater regional detail and longer periods.

\begin{figure}[h]
\includegraphics[width=\textwidth]{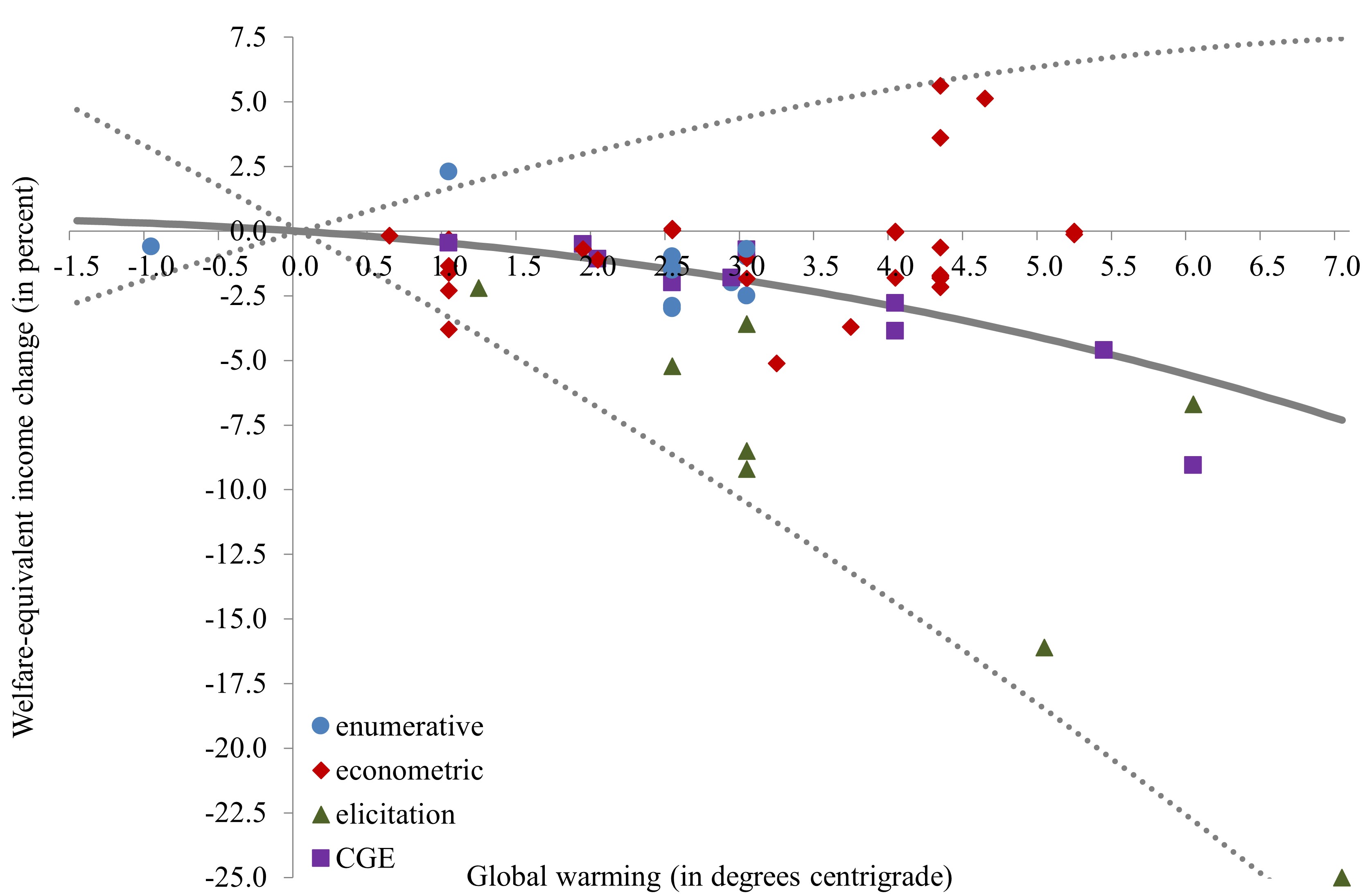}
\caption{The global total annual impact of climate change. Primary estimates are shown as dots. The central, solid line is the Bayesian model average, the dashed lines the 90\% confidence interval.}
\label{fig:totalimpact}
\end{figure}

For the moment, however, Figure \ref{fig:totalimpact} presents our best knowledge on the economic impacts of climate change. It is this information, warts and all, that should be used to estimate the social cost of carbon and inform the optimal rate of emission reduction\textemdash at least until new, hopefully better studies shine a different light on this question.

\bibliography{master}

\newpage
\begin{longtable}{|l |r |r |r |r |r| c|r|}
\caption{Estimates of the comparative static impact on global economic welfare}
\label{tab:compstat}
\footnotesize
\\ \hline \multicolumn{1}{|c|}{Study} & \multicolumn{1}{|c|}{warming} & \multicolumn{1}{|c|}{impact} & \multicolumn{1}{|c|}{st.dev.} & \multicolumn{1}{|c|}{min} & \multicolumn{1}{|c|}{max} & \multicolumn{1}{|c|}{method} & \multicolumn{1}{|c|}{scc} \\ \hline 
\endfirsthead

\multicolumn{8}{c}%
{{\bfseries \tablename\ \thetable{} -- continued from previous page}}
\\
\hline \multicolumn{1}{|c|}{Study} & \multicolumn{1}{|c|}{warming} & \multicolumn{1}{|c|}{impact} & \multicolumn{1}{|c|}{st.dev.} & \multicolumn{1}{|c|}{min} & \multicolumn{1}{|c|}{max} & \multicolumn{1}{|c|}{method} & \multicolumn{1}{|c|}{scc}\\ \hline 
\endhead

\hline \multicolumn{8}{|r|}{{Continued on next page}} \\ \hline
\endfoot 

\hline 
\endlastfoot

\citet{dArge1979}	&	-1.0	&	-0.6	&		&		&	& enum	& 92\\ \hline
\citet{Nordhaus1982}	&	2.5	&	-3.0	&		&	-12.0	&	5.0	& enum & 74\\ \hline
\citet{Nordhaus1991}	&	3.0	&	-1.0	&		&		&	& enum	& 17\\ \hline
\cite{Nordhaus1994bk}	&	3.0	&	-1.3	&		&		& & enum & 23		\\ \hline
\cite{Nordhaus1994as}	&	3.0	&	-3.6	&		&	-21.0	&	0.0 & elicit & 62	\\
	&	6.0	&	-6.7	&		&		&	& & 29	\\ \hline
\citet{Fankhauser1995}	&	2.5	&	-1.4	&		&		& & enum	& 35	\\ \hline
\citet{Berz}	&	2.5	&	-1.5	&		&		& & enum		& 37 \\ \hline
\citet{Schauer1995} & 2.5 & -5.22 & 8.44 & & & elicit & 129\\ \hline
\citet{Tol1995}	&	2.5	&	-1.9	&		&		& & enum	& 47	\\ \hline
\citet{NordhausYang1996}	&	2.5	&	-1.4	&		&		& & enum & 35		\\ \hline
\citet{PlambeckHope1996}	&	2.5	&	-2.9	&		&	-13.1	&	-0.5 & enum & 71	\\ \hline
\citet{Mendelsohn2000}	&	2.5	&	0.03	&	0.05	&		& & ectric	& -0.7	\\
	&	2.5	&	0.10	&	0.01	&		& &	& -2.5	\\
	&	4.0	&	-0.01	&		&		& & & 0.1		\\
	&	4.0	&	-0.04	&		&		& &	& 0.4	\\
	&	5.2	&	-0.01	&		&		& &	& 0.1	\\
	&	5.2	&	-0.13	&		&		& &	& 0.7	\\ \hline
\citet{NordhausBoyer2000} &	2.5	&	-1.5	&		&		&	& enum	& 37\\ \hline
\citet{Tol2002ERE1}	&	1.0	&	2.3	&	1.0	&		& & enum	& -355	\\ \hline
\citet{Maddison2003}	&	2.5	&	0.0	&		&		&	& ectric & -0.8	\\ \hline
\citet{Rehdanz2005}	&	0.6	&	-0.2	&		&		&	& ectric	& 77\\
	&	1.0	&	-0.3	&		&		& &	& 48	\\ \hline
\citet{Hope2006c} &	2.5	&	-1.0	&		&	-3.0	&	0.0	& enum & 24 \\ \hline
\citet{Nordhaus2006}	&	3.0	&	-0.9	&	0.1	&		& & ectric	& 16	\\
	&	3.0	&	-1.1	&	0.1	&		& & & 18		\\ \hline
\citet{Nordhaus2008}	&	3.0	&	-2.5	&		&		& & enum	& 43	\\ \hline
\citet{Horowitz2009} & 1.0 & 3.8 & & -4.2 & -2.7 & ectric & 587\\ \hline
\citet{Eboli2010} & 3.0 & -1.35 & & & & CGE & 23\\ \hline
\citet{Hope2011} & 3.0 & -0.7 & & -1.8 & -0.3 & enum & 12\\ \hline
\citet{Maddison2011}	&	3.2	&	-5.1	&		&		& & ectric	& 77	\\ \hline
\citet{Ng2011} & 1.0 & -1.35 & & & & ectric & 209 \\
& 1.0 & -1.61 & & & & & 249\\ \hline
\citet{Bosello2012}	&	1.9	&	-0.5	&		&		&	& CGE & 21	\\ \hline
\citet{Roson2012}	&	2.9	&	-1.8	&		&		&	& CGE & 33	\\
	&	5.4	&	-4.6	&		&		& & &  24		\\ \hline
\citet{McCallum2013} & 2 & -0.7 	&		&		&	& CGE & 27	\\
& 4 & -1.8 	&		&		&	&  & 17	\\ \hline
\citet{Nordhaus2013}	&	2.9	&	-2.0	&		&		&	& enum & 37	\\ \hline
\citet{Desmet2015}	&	4.6	&	5.1	&		&		&	& ectric & -37	\\
	&	9.3	&	-4.9	&		&		&	& & 9	\\
	&	13.6	&	-24.1	&		&		& & &	20	\\
	&	16.7	&	-78.9	&		&		&	& & 44	\\ \hline
\citet{Sartori2016}	&	3.0	&	-0.7	&		&		&	& CGE & 12	\\ \hline
\citet{Kompas2018}	&	1.0	&	-0.5	&		&		& & CGE	& 72	\\
	&	2.0	&	-1.1	&		&		&	& & 41	\\
	&	3.0	&	-1.8	&		&		&	& & 32	\\
	&	4.0	&	-2.8	&		&		&	& & 27	\\ \hline
\citet{Dellink2019}	&	2.5	&	-2.0	&		&		&	& CGE & 49	\\ \hline
\citet{Takakura2019}	&	2.0	&	-1.1	&	0.1	&		&	& CGE & 43	\\
	&	4.0	&	-3.9	&	0.4	&		&	& & 37	\\
	&	6.0	&	-9.1	&	1.6	&		&	& & 39	\\ \hline
\citet{Howard2020}	&	3.0	&	-9.2	&	10.3	&	-20	&	-2 & elicit	& 158 \\ \hline
\citet{Kalkuhl2020}	&	1	&	-2.3	&	1.32	&		&	& ectric & 355	\\ \hline
\citet{Conte2021}	&	3.7	&	-3.7	&		&		& & ectric & 42		\\ \hline
\citet{Cruz2021}	&	7.2	&	-5.0	&		&		& & ectric & 15		\\ \hline
\citet{Howard2021}	&	1.2	&	-2.2	&	2.9	&		&	& elicit	& 236\\
	&	3.0	&	-8.5	&	6.7	&		&	& & 146	\\
	&	5.0	&	-16.1	&	13.3	&		&	 & & 100	\\
	&	7.0	&	-25.0	&	20.7	&		&	& & 79	\\ \hline
\citet{Newell2021}	&	4.3	&	5.63	&		&		& & ectric & -47		\\
	&	4.3	&	3.61	&		&		&	&	& -30 \\
	&	4.3	&	-1.71	&		&		&	&	& 14\\
	&	4.3	&	-1.63	&		&		&	&	& 14\\
	&	4.3	&	-2.17	&		&		&	&	& 18\\
	&	4.3	&	-0.64	&		&		&	&	& 5\\
	&	4.3	&	-1.82	&		&		&	&	& 15\\
	&	4.3	&	-1.75	&		&		&	&	& 15\\
	&	4.3	&	-2.16	&		&		&	&	& 18\\ \hline
\multicolumn{8}{|l|}{\scriptsize \thead[l]{Valuation methods are \textbf{enum}erative, \textbf{elicit}ation, \textbf{ec}onome\textbf{tric}, and \textbf{c}omputable \textbf{g}eneral \textbf{e}quilibrium.\\The social cost of carbon, in 2010 US dollar per metric tonne of carbon, is for emissions in the year 2015, the SSP2 \\ scenario, a pure rate of time preference of 1\% and a rate of risk aversion of 1; impacts are proportional to temperature \\squared.}}   
\end{longtable}

\newpage
\begin{table}[]
    \centering \footnotesize
    \caption{Impact functions}
    \label{tab:function}
    \begin{tabular}{|l|c|c|l|r|} \hline
    name & function  & likelihood & proponent & scc\\ \hline
Parabolic & $-0.37T -0.098T^2 $ &	18.96\%  & \citet{Tol2009JEP} & 29\\
Hyperbolic sine & $\sfrac{1- e^{2 \cdot 0.40 T}}{2 e^{0.40 T}}$	& 18.36\% & this study & 22\\
Quadratic & $-0.17 T^2$ &	14.09\% & \citet{Nordhaus1992} & 27\\
Weitzman 6& $-0.18 T^2+ 7.38 \cdot 10^{-6} T^6$ &	13.59\% & \citet{Weitzman2012} & 29\\
Weitzman 7& $-0.18 T^2+ 9.85 \cdot 10^{-7} T^7$ &	13.49\%  & \citet{Weitzman2012}\footnote{Weitzman actually raises temperature to the power 6.754, which excludes cooling.} & 28\\
Linear & $-0.79T$ & 	11.50\%  & \citet{Hope2006c}\footnote{Hope is actually piecewise linear, with zero impacts below one temperature threshold, linear damages above that threshold, and shifted linear damages above another temperature threshold.} & 29\\
Piecewise linear & \thead{$-0.79T \text{ if } T \leq 12.8$ \\ $-0.79 \cdot 12.8 - 17.8(T-12.8)  \text{ if }  T > 12.8$} &	9.78\%  & \citet{Tol2018} & 29\\
Exponential & $0.0077 - 0.0077e^T$ &	0.22\% & \citet{Ploeg2014EER} & 3 \\  \hline
\end{tabular}
\caption*{\scriptsize The social cost of carbon, in 2010 US dollar per metric tonne of carbon, is for emissions in the year 2015, the SSP2 scenario, a pure rate of time preference of 1\%, and a rate of risk aversion of 1. }   
\end{table}

\begin{table}[]
\footnotesize
    \centering
        \caption{Sectoral impacts of 2.5\celsius{} global warming. Impacts are expressed as a percentage of gross global income. Imputed impacts are in \textit{italics}.}
    \label{tab:sector}
    \begin{tabular}{|l|r|r|r|r|r|r|r|r|r|} \hline
	&	Fankhauser	&	Berz	&	Tol	&	Nordhaus	&	Bosello	&	Dellink	&	Sartori	&	Kompas	&	Average	\\ \hline
Agriculture	&	-0.20	&	-0.19	&	-0.13	&	-0.13	&	-0.40	&	-0.75	&	-0.10	&	-0.22	&	-0.26	\\
Forestry	&	-0.01	&	-0.02	&	\textit{-0.01}	&	0.00	&	-0.01	&	\textit{-0.01}	&	\textit{-0.01}	&	-0.02	&	-0.01	\\
Energy	&	-0.12	&	-0.11	&	\textit{-0.07}	&	-0.02	&	-0.02	&	-0.06	&	\textit{-0.07}	&	-0.06	&	-0.07	\\
Water	&	-0.24	&	-0.23	&	\textit{-0.10}	&	-0.03	&	-0.01	&	\textit{-0.10}	&	\textit{-0.10}	&	-0.01	&	-0.10	\\
Tourism	&	\textit{-0.09}	&	\textit{-0.09}	&	\textit{-0.09}	&	\textit{-0.09}	&	-0.17	&	-0.14	&	0.04	&	-0.10	&	-0.09	\\
Other markets	&	\textit{-0.63}	&	\textit{-0.63}	&	\textit{-0.63}	&	\textit{-0.63}	&	\textit{-0.63}	&	\textit{-0.63}	&	-0.29	&	-0.98	&	-0.63	\\
Coastal defence	&	0.00	&	-0.01	&	-0.08	&	-0.04	&	-0.03	&	-0.04	&	0.00	&	-0.02	&	-0.03	\\
Dryland loss	&	-0.07	&	-0.07	&	-0.09	&	-0.09	&	-0.08	&	-0.10	&	0.00	&	-0.07	&	-0.07	\\
Wetland loss	&	-0.16	&	-0.16	&	-0.17	&	-0.19	&	\textit{-0.17}	&	\textit{-0.17}	&	\textit{-0.17}	&	\textit{-0.17}	&	-0.17	\\
Ecosystem	&	-0.21	&	-0.20	&	-0.19	&	-0.20	&	\textit{-0.20}	&	\textit{-0.20}	&	\textit{-0.20}	&	\textit{-0.20}	&	-0.20	\\
Health	&	-0.26	&	-0.40	&	-0.77	&	-0.10	&	-0.01	&	-0.90	&	-0.19	&	-0.38	&	-0.38	\\
Air pollution	&	-0.08	&	-0.08	&	\textit{-0.08}	&	\textit{-0.08}	&	\textit{-0.08}	&	\textit{-0.08}	&	\textit{-0.08}	&	\textit{-0.08}	&	-0.08	\\
Time use	&	\textit{0.29}	&	\textit{0.29}	&	\textit{0.29}	&	0.29	&	\textit{0.29}	&	\textit{0.29}	&	\textit{0.29}	&	\textit{0.29}	&	0.29	\\
Settlements	&	\textit{-0.17}	&	\textit{-0.17}	&	\textit{-0.17}	&	-0.17	&	\textit{-0.17}	&	\textit{-0.17}	&	\textit{-0.17}	&	\textit{-0.17}	&	-0.17	\\
Catastrophe	&	-0.01	&	-0.01	&	-0.01	&	-1.02	&	\textit{-0.27}	&	\textit{-0.27}	&	\textit{-0.27}	&	\textit{-0.27}	&	-0.27	\\
Migration	&	-0.02	&	-0.02	&	-0.12	&	\textit{-0.06} 	& \textit{-0.06}	&	\textit{-0.06}	&	\textit{-0.06}	&	\textit{-0.06}	& -0.06\\
Amenity	&	\textit{-0.33}	&	\textit{-0.33}	& -0.33	&	\textit{-0.33}	&	\textit{-0.33}	&	\textit{-0.33}	&	\textit{-0.33}	&	\textit{-0.33}	&	-0.33	\\
Total	&	-2.34	&	-2.44	&	-2.77	&	-2.90	&	-2.18	&	-3.56	&	-1.54	&	-2.77	&	-2.56	\\
Original	&	-1.40	&	-1.50	&	-1.90	&	-1.50	&	-0.50	&	-2.00	&	-0.55	&	-1.40	&	-1.58	\\
Ratio	&	1.67	&	1.63	&	1.46	&	1.93	&	2.98	&	1.78	&	2.82	&	1.98	&	1.63	\\ \hline
    \end{tabular}
\end{table}

\newpage

\appendix

\setcounter{table}{0}
\renewcommand{\thetable}{B\arabic{table}}
\setcounter{figure}{0}
\renewcommand{\thefigure}{B\arabic{figure}}

\section{Differences with Howard and Sterner (2017)}
\label{sc:howard}
\citet{Howard2017} include a number of estimates that I excluded, for the following reasons. The paper by \citet{MeyerCooper1995} is confused, as explained in \citet{FankhauserTol1995}. \citet{Howard2017} report an impact of 11.5\% of GDP for 3.0\celsius{} warming, a number that is not in \citet{MeyerCooper1995}. \citet{Hanemann2008} presents estimates for the USA only, \citet{Bluedorn2009} present no estimate. \citet{Manne1995} do not present new impact estimates, instead rely on Nordhaus and Fankhauser. \citet{Howard2017} misinterpret their estimate, an annuity, as a point on the impact function. \citet{Ackerman2012, Bosetti2007, Gunasekera2008, ManneRichels2005, Tol2013CC, Weitzman2012} do not present \emph{new} estimates of the total impact of climate change. \citet{Anthoff2022} count over 200 papers that estimate the social cost of carbon, each of which has at least one estimate of the total impact of climate change\textemdash but very few of these papers present \emph{new} estimates of the total impact. It is not clear why \citet{Howard2017} included 6 of these papers but not the other 194 or so.

\section{Additional results}
\begin{figure}[h]
\includegraphics[width=\textwidth]{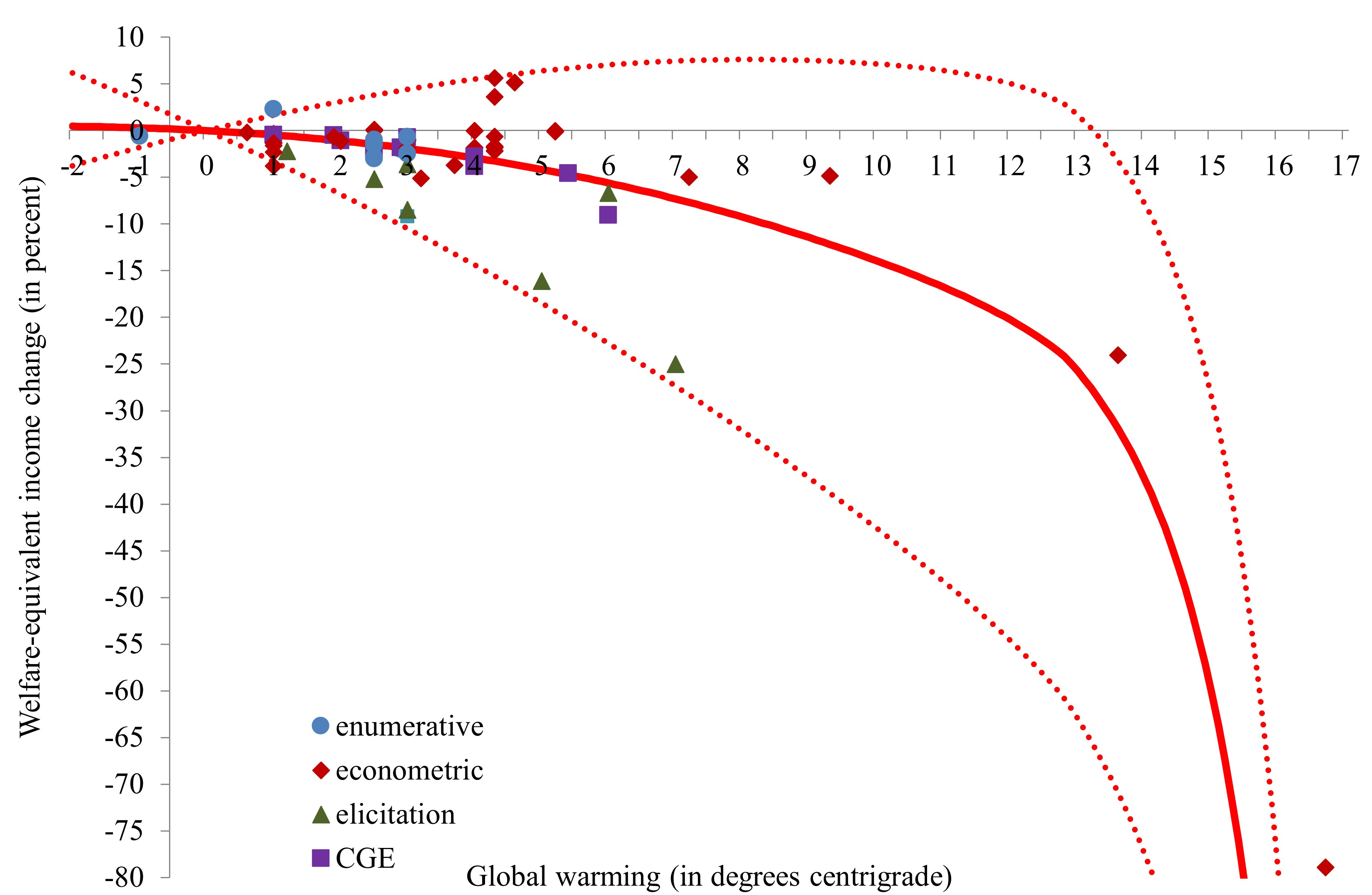}
\caption{The global total annual impact of climate change. Primary estimates are shown as dots. The central, solid line is the Bayesian model average, the dashed lines the 90\% confidence interval.}
\label{fig:totalimpactall}
\end{figure}

\begin{table}[]
    \centering
     \caption{Relative likelihood of the impact function by primary estimate method. The best-fitting curve is marked in \textbf{bold}.}
    \label{tab:methods}
       \begin{tabular}{l r r r r r} \hline
function	&	all	&	enumerative	&	elicitation	&	econometric	&	CGE	\\ \hline
Parabolic	&	\textbf{20.7}\%	&	6.4\%	&	20.0\%	&	10.8\%	&	9.6\%	\\
Hyperbolic sine	&	17.6\%	&	3.0\%	&	9.0\%	&	12.6\%	&	14.5\%	\\
Linear	&	14.0\%	&	2.6\%	&	\textbf{26.9}\%	&	11.5\%	&	0.2\%	\\
Piecewise linear	&	13.3\%	&	\textbf{41.9}\%	&	8.2\%	&	14.3\%	&	13.2\%	\\
Weitzman 6	&	11.6\%	&	17.9\%	&	10.0\%	&	12.8\%	&	24.5\%	\\
Weitzman 7	&	11.5\%	&	19.5\%	&	9.6\%	&	13.3\%	&	\textbf{28.4}\%	\\
Quadratic	&	11.2\%	&	6.2\%	&	12.7\%	&	10.4\%	&	8.8\%	\\
Exponential	&	0.1\%	&	2.5\%	&	3.7\%	&	\textbf{14.3}\%	&	0.74\%	\\ \hline
SCC & 27 & 57 & 87 & 1 & 6 \\ \hline
\end{tabular}

\caption*{\scriptsize The social cost of carbon, in 2010 US dollar per metric tonne of carbon, is for emissions in the year 2015, the SSP2 scenario, a pure rate of time preference of 1\%, and a rate of risk aversion of 1. The function used is the one given in bold, except for the column ``all'', where the weighted average is used.}   
\end{table}

\begin{table}[]
    \centering
    \caption{Fitted global impact function for the studies that relate economic growth to the temperature level.}
    \label{tab:level}
\begin{tabular}{l r r r}\hline
study & $T$ & $T^2$ & scc \\ \hline
\citet{Dell2012} & -0.002131 & 0 & 222 \\
\citet{Burke2015} & 0.001558 & -0.000500 & 33\\
\citet{Pretis2018} & 0.002586 & -0.000400 & -171\\
\citet{Henseler2019} & -0.000810 & -0.000455 & 259\\
\citet{Acevedo2020} & 0.002105 & -0.000510 & -37 \\
\citet{Kalkuhl2020} &-0.006666 & -0.000015 & 463\\
\citet{Kikstra2021} & -0.010098 & -0.000761 & 539\\
Combined & -0.002023 & -0.000023 & 218 \\ \hline
\end{tabular}
\caption*{\scriptsize The social cost of carbon, in 2010 US dollar per metric tonne of carbon, is for emissions in the year 2015, the SSP2 scenario, a pure rate of time preference of 1\%, and a rate of risk aversion of 1. }   
\end{table}

\begin{table}[]
    \centering
    \caption{Estimated global impact function for the studies that relate economic growth to the temperature change.}
    \label{tab:change}
\begin{tabular}{l r r r}\hline
study & $\Delta T$ & $T \Delta T$ & scc \\ \hline
\citet{Letta2018} & -0.004655 & 0 & 16 \\
\citet{Kahn2019} & 0.01714 & 0 & 60\\
\citet{Kotz2021} & 0.0019 & -0.0018 & 17\\
\citet{Kotz2022} &-0.0013 & -0.0018 & 18\\
Combined & -0.004029 & -0.000377 & 17 \\ \hline
\end{tabular}
\caption*{\scriptsize The social cost of carbon, in 2010 US dollar per metric tonne of carbon, is for emissions in the year 2015, the SSP2 scenario, a pure rate of time preference of 1\%, and a rate of risk aversion of 1. }   
\end{table}

\begin{figure}[h]
\includegraphics[width=\textwidth]{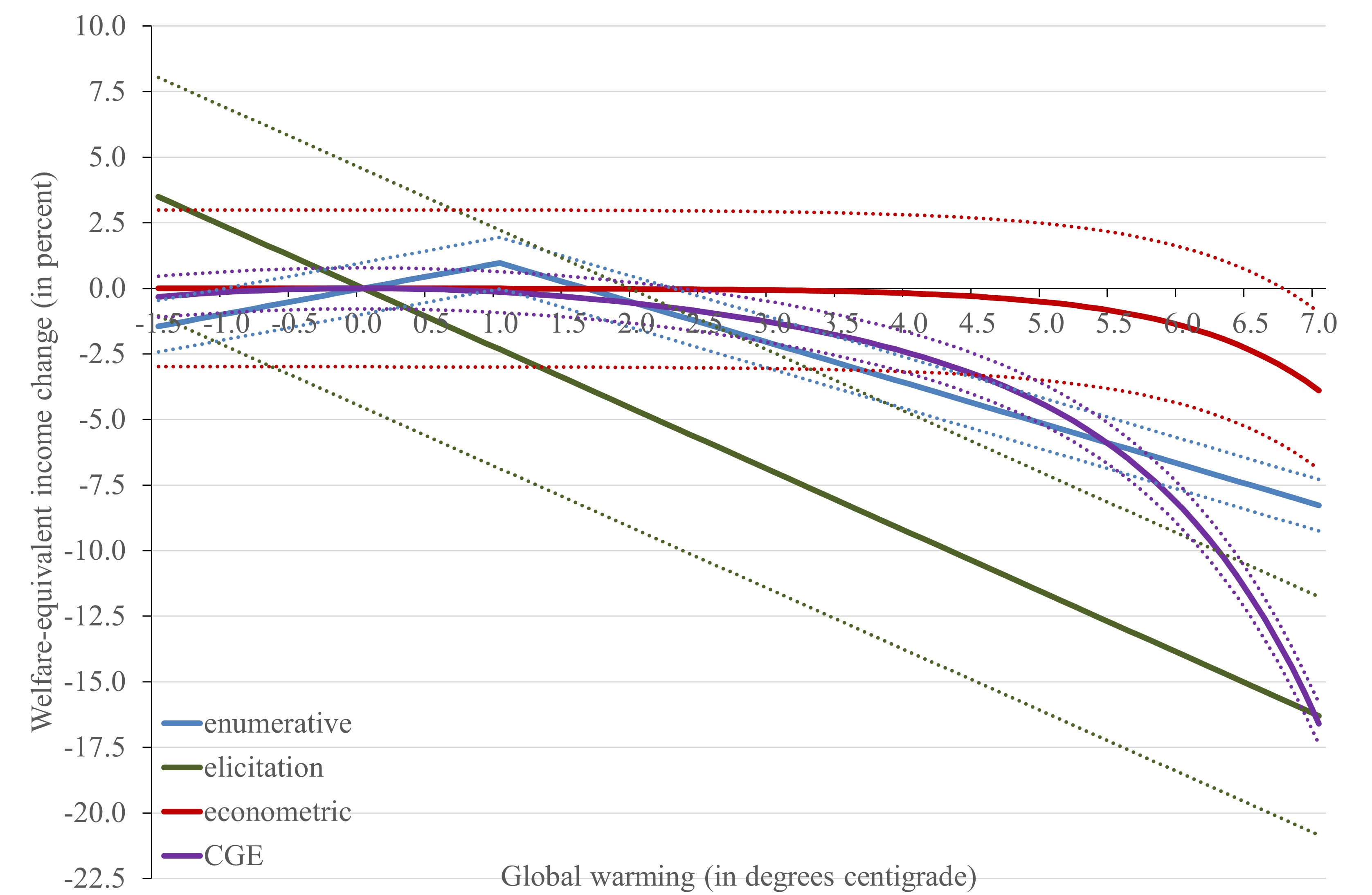}
\caption{Best model fits for the global total annual impact of climate change. The central, solid line is the central estimate, the dashed lines plus or minus the standard error of regression.}
\label{fig:method}
\end{figure}

\end{document}